\documentclass[
 reprint,
 amsmath,amssymb,
 aps,
 prb
]{revtex4-2}

\usepackage{graphicx}
\usepackage{dcolumn}
\usepackage{bm}
\usepackage[colorlinks=true,linkcolor=blue,allcolors=blue]{hyperref}
\usepackage{color}
\usepackage[OT1]{fontenc}

\begin{document}

\preprint{prb}

\title{Nonequilibrium Statistics of Biased Kondo Resonance}
\author{Jong E. Han}
\email{jonghan@buffalo.edu}
\affiliation{Department of Physics, State University of New York at Buffalo, Buffalo, New York 14260, USA}

\date{\today}

\begin{abstract} 

Numerical renormalization group (NRG) is formulated for nonequilibrium steady-state by converting finite-lattice many-body eigenstates into scattering states. Extension of the full-density-matrix NRG for a biased Anderson impurity model, \textcolor{black}{simplified by formulating with the original orbital basis as the Hamiltonian,} enables detailed studies of the sub-Kondo spectral evolution in the zero-temperature limit, \textcolor{black}{confirming} the double-resonance structure at bias of the Kondo energy scale $T_K$. \textcolor{black}{The distribution shows distinct multi-scale spectral features at energy $\omega$ below the Kondo scale ($\omega\lesssim T_K$) and near the bias ($\omega\gtrsim V$), leading to the nonequilibrium temperature $T_{\rm loc}$ local to the Kondo dot scaling as $k_BT_{\rm loc}\approx V$ for $V\gg T_K$. The current-voltage relation in the low-temperature limit ($T\ll T_K$) deviates from the unitary limit as the bias exceeds the Kondo scale ($V/2\gtrsim T_K$) and reaches the current saturation regime.}

\end{abstract}

\maketitle

\section{Introduction}

\textcolor{black}{Solving nonequilibrium statistical mechanics problems} in biased quantum systems
has been a tremendous challenge in recent decades. In nonequilibrium quantum many-body theories, much of the focus has been placed on quantum impurity models~\cite{KondoPTP1964,hewson1997kondo,doyonPRB2006,mehtaPRL2006,meirPRL1992,uedaPRB2003,ErpenbeckPRL2023,hanPRL2007,arrigoniPRL2013,dordaPRB2014,andersPRL2008} as the prototype strong correlation model and its pivotal role for solving the condensed matter limit through the dynamical mean-field theory~\cite{GeorgesRMP1996,AokiRMP2014}. Despite the huge effort, the theoretical rigor and the numerical efficiency of the nonequilibrium impurity solvers have yet to reach a satisfactory level at which the techniques are routinely applicable to nonequilibrium lattice calculations as in equilibrium lattice problems.

The Kondo problem~\cite{KondoPTP1964,hewson1997kondo} remains the testbed of numerous strongly correlated nonequilibrium techniques such as  analytical theories~\cite{doyonPRB2006,mehtaPRL2006,schillerPRB1995,BolechPRB2016}, diagrammatic methods~\cite{uedaPRB2003,meirPRL1992,oguriJPSJ2005}, real-time (RT) diagrammatic simulation~\cite{ErpenbeckPRL2023}, imaginary-time simulation~\cite{hanPRL2007,hanPRB2010}, numerical renormalization group (NRG)~\cite{andersPRL2008,schwarzPRL2018}, \textcolor{black}{RT renormalization group (RG)~\cite{SchoellerRTRG}, perturbative RG~\cite{RoschPRL2003}, functional RG~\cite{GezziPRB2007}, RT path-integral~\cite{WeissRTPI}, time-dependent density-matrix RG~\cite{daSilva,Heidrich-Meisner_PRB}, tensor-network~\cite{Wauters,jeannin2025}}, and auxiliary density-matrix~\cite{arrigoniPRL2013,dordaPRB2014} methods. The debate on how the Kondo resonance dissipates under bias, namely whether the main Kondo peak splits into a double-peak with its spacing equal to the bias, has largely been settled. However, computational difficulties have limited the understanding of what happens to the Kondo singlet when it breaks into a double-peak with its energy scale in the sub-Kondo regime. \textcolor{black}{Furthermore, compared to the spectral features in the model, the nonequilibrium statistics of the spectra has not been understood as much}. It is the purpose of this work to provide clarity by building on the well-established NRG method with accessible extensions for steady-state nonequilibrium.

The model of an Anderson impurity coupled to the source (or left, $L$) and drain (or right, $R$) reservoirs can be written as the Hamiltonian
\begin{eqnarray}
\hat{H} & = & \sum_\sigma \epsilon_d n_{d\sigma} + U\left(n_{d\uparrow}-\frac12\right)\left(n_{d\downarrow}-\frac12\right)
\label{eq:ham} \\
& & + \sum_{\alpha k\sigma}\epsilon_k c^\dagger_{\alpha k\sigma}c_{\alpha k\sigma} -\frac{t}{\sqrt{\Omega}}
\sum_{\alpha k\sigma}(d^\dagger_\sigma c_{\alpha k\sigma}+c^\dagger_{\alpha k\sigma}d_\sigma),\nonumber
\end{eqnarray}
with the creation/annihilation operators $d^\dagger_\sigma$/$d_\sigma$ for the impurity $d$-orbital of spin $\sigma=\uparrow,\downarrow$, and $c^\dagger_{\alpha k\sigma}$/$c_{\alpha k\sigma}$ for the conduction electrons with the continuum index $k$ and the reservoir index $\alpha=L,R$. The impurity occupation number (per spin) is $n_{d\sigma}=d^\dagger_\sigma d_\sigma$, $U$ the Coulomb repulsion parameter and $\epsilon_d$ the impurity level-energy. \textcolor{black}{With the Coulomb term written in the particle-hole (p-h) symmetric form, the p-h symmetric limit is $\epsilon_d=0$.} The continuum energy for the reservoirs $\epsilon_k$ is assumed to be independent of $\alpha$ and $\sigma$, for simplicity. We further assume that $\epsilon_k$ forms a flat density of states with the (half) bandwidth $D$. We set $D=1$ as the unit of energy. The mixing of $d$-orbital with the conduction band, given as $t$ (with the factor $\Omega$ \textcolor{black}{for normalization by the reservoir volume}), is parametrized by the hybridization $\Gamma_\alpha=\pi t^2\Omega^{-1}\sum_k\delta(\epsilon_k)$. In this work, we focus on the symmetric coupling $\Gamma_L=\Gamma_R$ and define the total hybridization as $\Gamma=\Gamma_L+\Gamma_R$. We use the convention $\hbar=e=k_B=1$ with the Planck constant $\hbar$, electric charge $e$, and the Boltzmann constant $k_B$.

The steady-state nonequilibrium density-matrix $\hat{\rho}_{\rm neq}$ driven by two chemical potentials, $\mu_L=V/2$ and $\mu_R=-V/2$ for $L$ and $R$-lead respectively, is constructed as
\begin{equation}
\hat{\rho}_{\rm neq}=\frac{1}{Z_{\rm neq}}e^{-\beta(\hat{H}-V\hat{Y})} \mbox{ with }\beta=1/T,
\label{eq:rho}
\end{equation}
where Hershfield's $Y$-operator~\cite{hershfield1993}
\begin{equation}
\hat{Y}=\frac12\sum_{k\sigma}({\psi^\dagger_{Lk\sigma}\psi_{Lk\sigma}-\psi^\dagger_{Rk\sigma}\psi_{Rk\sigma}})
\end{equation}
is constructed from the full scattering-state operators $\psi^\dagger_{\alpha k\sigma}$ as solutions to the Hamiltonian $\hat{H}$ through the Lippmann-Schwinger equation~\cite{merzbacher,gellmann1953,hanPRL2007,hanPRB2010}. $Z_{\rm neq}={\rm Tr}e^{-\beta(\hat{H}-V\hat{Y})}$ is the normalization so that ${\rm Tr\,\hat{\rho}_{\rm neq}}=1$. With $\psi^\dagger_{\alpha k\sigma}$ being energy eigen-operators of $\hat{H}$, the density-matrix $\hat{\rho}_{\rm neq}$ is time-independent with the multiple chemical potentials defined separately in each reservoir. Despite its deceptively simple form, application of the formulation has been much limited since its construction requires the knowledge of the full eigen-operators. \textcolor{black}{This work rests on the idea that the NRG method~\cite{wilsonRMP1975,KWWPRB1980,bullaRMP2008,andersPRL2008} is particularly well-suited to realize the idea since it iteratively constructs the low-energy many-body eigenstates explicitly.}  

\begin{figure}
\begin{center}
\rotatebox{0}{\resizebox{3.4in}{!}{\includegraphics{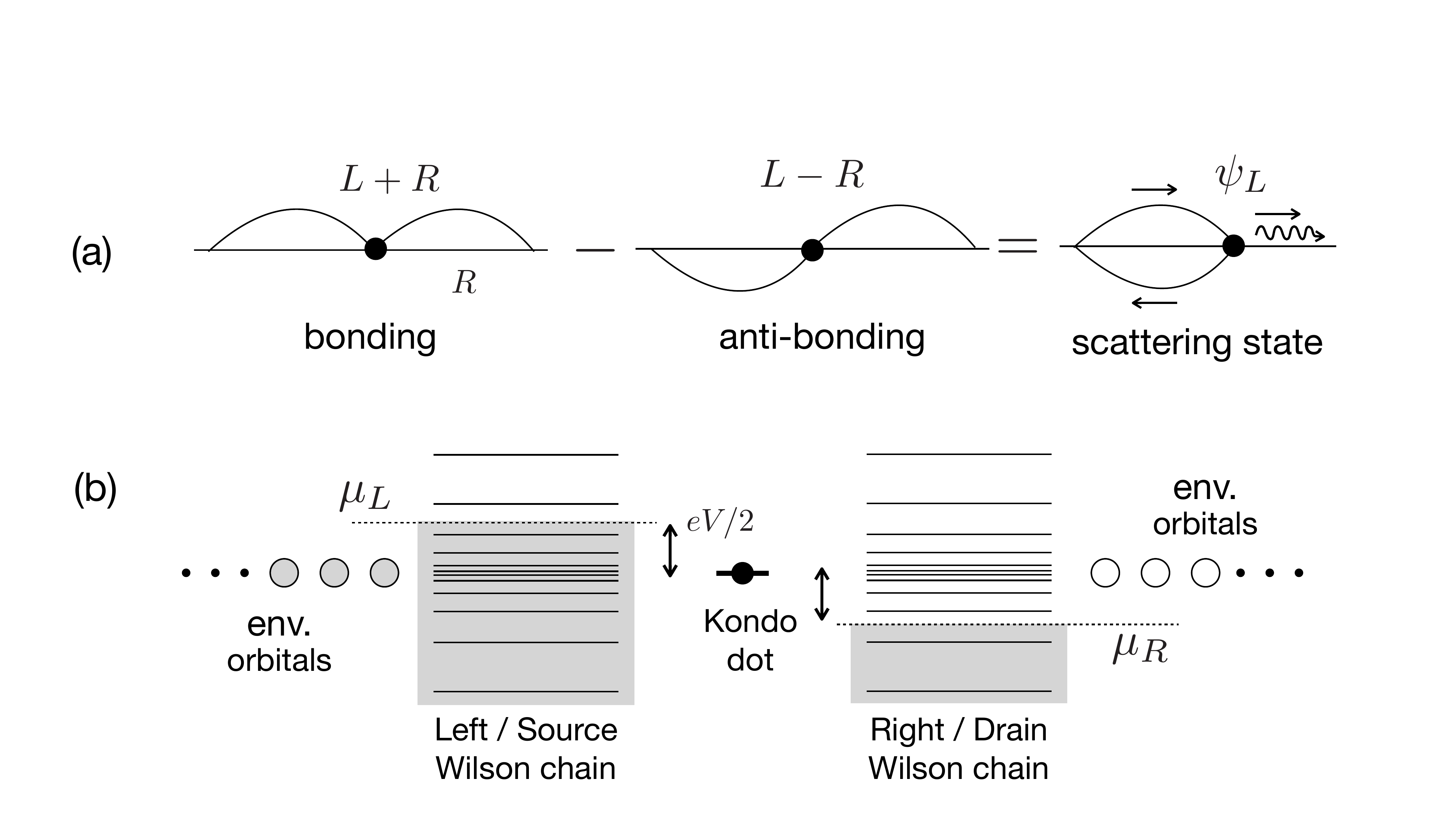}}}
\caption{\textcolor{black}{(a) Scattering state represented as a linear combination of bonding and anti-bonding states computed on finite chains. By choosing a certain linear superposition of the bonding and anti-bonding states, the wave $\psi_L$ can be made to have only the right-moving component in the $R$-reservoir.} (b) The nonequilibrium NRG scheme for an Anderson impurity~\cite{hewson1997kondo} coupled to biased reservoirs. The $L/R$ reservoir is made up of the Wilson chain of length $N$ with the $N_{\rm max}-N$ environment orbitals at energy levels at zero. The chemical potentials apply up to $\pm V/2$ to $L/R$-reservoirs, respectively, including the environment orbitals. The levels from the Wilson chains follow the conventional scaling of the Wilson tight-binding parameters~\cite{wilsonRMP1975}.
}
\label{fig1}
\end{center}
\end{figure}

\section{Formulation}

Before implementing the NRG for the nonequilibrium steady-state, we first discuss how finite-lattice eigenstates can be used for representing scattering states. As illustrated in Fig.~\ref{fig1}(a), eigenstates for a system made up of two leads (depicted as lines) and a quantum dot (black dot) can be understood as closely lying bonding and anti-bonding states in the tunneling window. \textcolor{black}{By forming a certain linear combination between them, we can create a right- or left-moving scattering state; For instance, by canceling the left-moving wave in the $R$-reservoir, we have a scattering state out-going from $L$, $\psi_L$, as depicted in (a).} While this is not an exact eigenstate, we take that as an approximate eigenstate of an open system. \textcolor{black}{As we approach the low energy scale with increasing WC length, the mapping to the scattering states becomes more robust.} 

The explicit construction of the scattering state operators follows the general scattering theory~\cite{gellmann1953},
\begin{equation}
    \hat{Y}=\frac{i\eta}{-{\cal L}+i\eta}\hat{Y}_0\mbox{ with }
    \hat{Y}_0=\frac12(\hat{N}_L-\hat{N}_R)
    \label{eq:y}
\end{equation}
where $\hat{Y}_0$, the initial $Y$-operator, is that of disconnected reservoirs from the impurity [i.e., $t=0$ in Eq.~(\ref{eq:ham})] with $\hat{N}_\alpha=\sum_{k\sigma}c^\dagger_{\alpha k\sigma}c_{\alpha k\sigma}$. Therefore, $\hat{Y}_0$ is trivially constructed from the WC under any bias. The Liouvillian ${\cal L}$ acting on the operator space is defined as ${\cal L}\hat{A}=[\hat{H},\hat{A}]$ for an arbitrary operator $\hat{A}$. With energy eigenstates $|\alpha\rangle$ and $|\beta\rangle$ of the total $\hat{H}$, the full $Y$-operator can be written as
\begin{equation}
\langle\alpha|\hat{Y}|\beta\rangle=\frac{i\eta}{-E_\alpha+E_\beta+i\eta}\langle\alpha|\hat{Y}_0|\beta\rangle.
\label{eq:yab}
\end{equation}
It is immediately clear that $\hat{Y}$ is Hermitian with ${\cal L}^\dagger=-{\cal L}$. The infinitesimal $\eta$ \textcolor{black}{term} sets the time scale $\eta^{-1}$ over which a scattering state is established~\cite{gellmann1953}. The sign of $\eta$ is set to represent out-going operators.

The infinitesimal $\eta$ parameter, set as finite during numerical calculations, plays two crucial roles: (1) $\eta$ anneals any operator into an eigen-operator of the full Hamiltonian $\hat{H}$ within the energy shell set by $\eta$, (2) Within the $\eta$-shell, it mixes eigenstates of a finite-lattice into approximate scattering states as depicted in Fig.~\ref{fig1}(a). Therefore, $\eta$ should be chosen as small but finite enough to mix the bonding and anti-bonding states. \textcolor{black}{Explicit determination of $\eta$ is discussed below.}

\textcolor{black}{Here, we briefly compare the method with the steady-state NRG (SNRG) method proposed by Anders~\cite{andersPRL2008}. The two methods, both founded on Hershfield's steady-state  formulation~\cite{hershfield1993}, differ mainly in the way that $\hat{Y}_0$ is constructed. The SNRG builds $\hat{Y}_0$ in terms of the non-interacting scattering states from both reservoirs. The basis from each reservoirs become entangled with the $d$-orbital, and the problem is transformed to an inter-site coupled system. On the contrary, the method proposed here uses the original basis of $\hat{H}$ and, therefore, the implementation is more straightforward and suited for extensions to generalized problems of complex impurity interactions and to future applications to nonequilibrium dynamical mean-field theory. The SNRG uses an explicit time-evolution for the full $\hat{Y}$ while an equivalent effect is achieved here by Eq.~(\ref{eq:y}) to obtain the density-matrix without an explicit construction of scattering states.}

\begin{figure}
\begin{center}
\rotatebox{0}{\resizebox{3.4in}{!}{\includegraphics{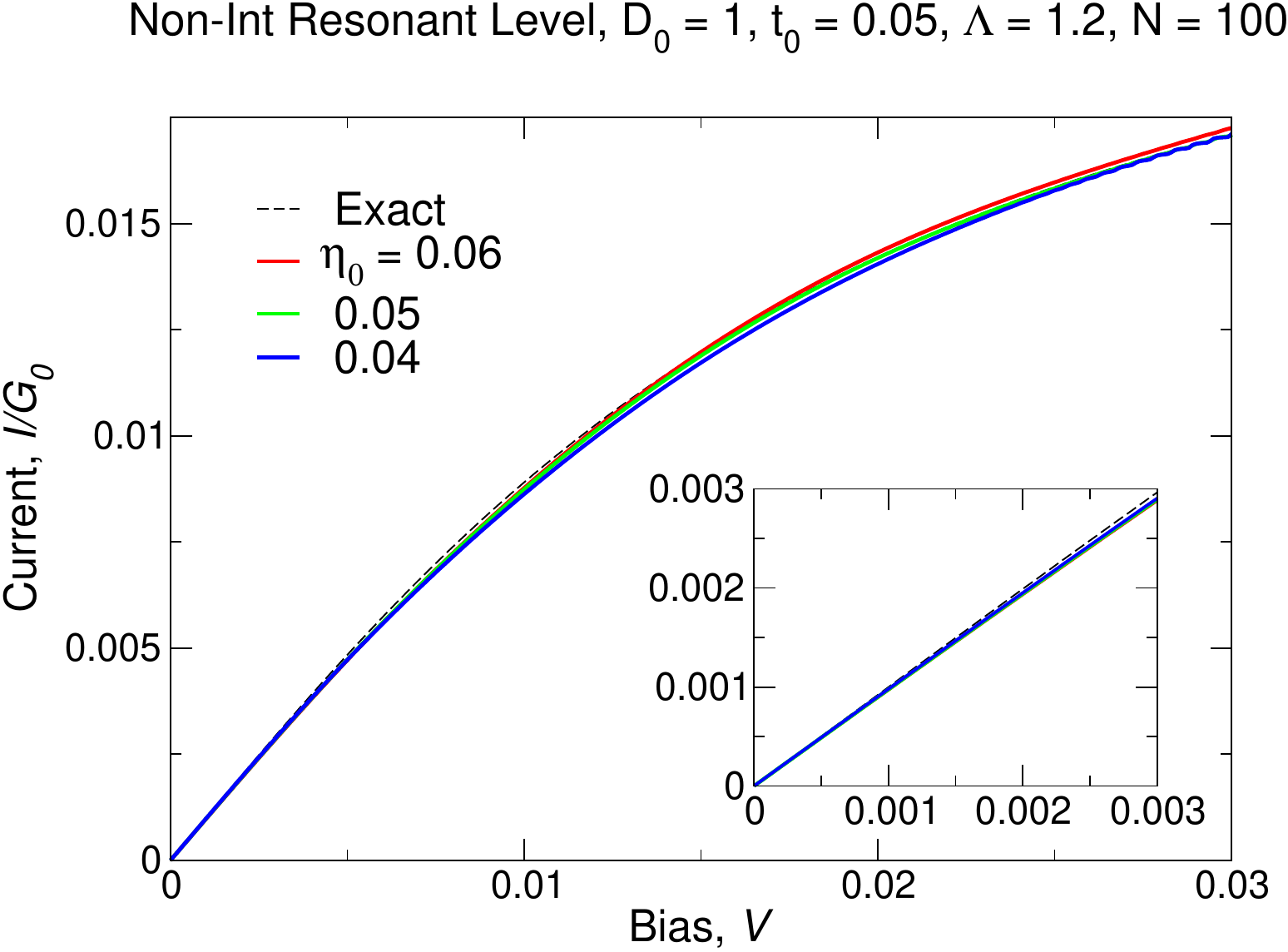}}}
\caption{Current calculated in the non-interacting resonant level model. The current expectation value $I$ (per spin, see text for definition) scaled with the conductance quantum $G_0=e^2/(2\pi\hbar)$ follows the exact result (dashed line) in the low bias limit (blown up in the inset). The infinitesimal parameter $\eta$ of Eq.~(\ref{eq:y}) is varied \textcolor{black}{below the hybridization $\Gamma=2\pi t^2/(2D)=0.0079$ with the hopping $t=0.05$ and the (half) bandwidth $D=1$. (See main text for the definition of $\eta_0$.} The level energy is set to the particle-hole symmetric limit $\epsilon_d=0$. \textcolor{black}{The $z$-averaging~\cite{oliveiraPRB1994} was used over 8 values of $z$ ($N_z=8$).}
}
\label{fig2}
\end{center}
\end{figure}

To test the concept of Eq.~(\ref{eq:y}) for DC transport, we perform an elementary calculation of the non-interacting resonant-level model [Eq.~(\ref{eq:ham}) with $U=0$] where the states $|\alpha\rangle$ and $|\beta\rangle$ in Eq.~(\ref{eq:yab}) can be represented by single-particle basis. We model the reservoirs by the WC~\cite{wilsonRMP1975,KWWPRB1980,bullaRMP2008} to densely sample the low-lying states as depicted in Fig.~\ref{fig1}(b). The $L/R$ reservoirs are modeled with respective WC having the number of sites $N_{WC}=100$, and the renormalization parameter $\Lambda=1.2$. The WCs can be easily diagonalized in the single-particle basis, and through Eqs.~(\ref{eq:rho}) and (\ref{eq:yab}), the nonequilibrium density-matrix is set up straightforwardly, and we may evaluate the current expectation value by using any single-particle theory. \textcolor{black}{Due to the logarithmic energy spacing of the WC, high bias calculations sample increasingly sparce levels. Therefore, we used the bias-dependent $\eta$-parametrization as $\eta={\rm min}(\eta_0\sqrt{\Gamma^2+V^2},\Gamma)$ that increases with $V$ but is bounded by the level-broadening $\Gamma$.}

The comparison of the current vs bias between the numerical and the exact~\cite{meirPRL1992} (dashed line) results is shown in Fig.~\ref{fig2}. The current is defined as $\langle I\rangle = it/(2\sqrt{\Omega})\sum_k\,{\rm Tr}\,[\hat{\rho}_{\rm neq}(c^\dagger_{Rk\sigma}d_\sigma-c^\dagger_{Lk\sigma}d_\sigma-H.C.)]$. Following the steps of Eqs.~(\ref{eq:rho}-\ref{eq:yab}), one can evaluate any steady-state observables. The transport at small bias agrees well with the linear response limit for the current (per spin) $I=G_0V$ with the unitary quantum conductance $G_0=1/(2\pi)$. \textcolor{black}{At large bias, numerical results follow the exact relation $I=(\Gamma/\pi)\tan^{-1}(V/2\Gamma)$. This demonstration confirms that the nonequilibrium density-matrix can be formulated with Hershfield's~\cite{hershfield1993} $Y$-operator, Eq.~(\ref{eq:y})~\cite{may_thesis}. To reduce the discreteness of the WC, we performed the $z$-averaging~\cite{oliveiraPRB1994} with 8 $z$-values. The code that produced Fig.~\ref{fig2} (without the $z$-averaging) is provided in Supplementary Material (SM)~\cite{SM}.}

\section{Results and Discussion}

We now turn to the interacting limit solved by NRG. In the interacting limit $U>0$, we use the full-density-matrix NRG (FDM-NRG)~\cite{andersschillerPRL2005,weichselbaumPRL2007} to incorporate the nonequilibrium excitations. As depicted in Fig.~\ref{fig1}(b), the reservoir is made up of the WC and the environment orbitals. 
The FDM-NRG uses the complete set of the Hilbert space of the WC of maximum length $N_{\rm max}$ by keeping record of the NRG-truncated states. The formulation of the FDM-NRG remains unchanged except that the steady-state density-matrix Eq.~(\ref{eq:rho}) is non-diagonal between the eigenstates of $\hat{H}$ within the same energy shell of width $\eta$. The algorithm for the reduced density-matrix~\cite{hofstetter2000,weichselbaumPRL2007} remains the same as that of the equilibrium FDM-NRG. 

One simple, yet crucially different step from equilibrium is the proper consideration of the free-energy from the environment orbitals under bias. The environment orbitals, despite being disconnected from the WC, are also subject to the bias as a part of the total system of length $N_{\rm max}$ and the increase of the WC length (i.e. the decrease of the environmental orbitals) by one site leads to the partition function $Z_{\rm env}=[(1+e^{V/2T})(1+e^{-V/2T})]^2$ and the free-energy change from the environment variables
\begin{equation}
    \Delta F_{\rm env}=V+4T\ln(1+e^{-V/2T}).
\end{equation}
The factor $e^{-\Delta F_{\rm env}/T}$ is multiplied to the density-matrix Eq.~(\ref{eq:y}) at each NRG iteration. The configurational weight $w_n$ in the FDM thus obtained peaks sharply~\cite{weichselbaumPRL2007} at the WC length $n=M$ reasonably well predicted by the condition of the NRG energy scale $D_n$,
\begin{equation}
    D_M\approx V/2\mbox{ with }D_n=\Lambda^{-(n-1)/2},
\label{eq:dn}
\end{equation}
with the NRG renormalization parameter $\Lambda$ (See SM~\cite{SM}).

\begin{figure}
\begin{center}
\rotatebox{0}{\resizebox{3.4in}{!}{\includegraphics{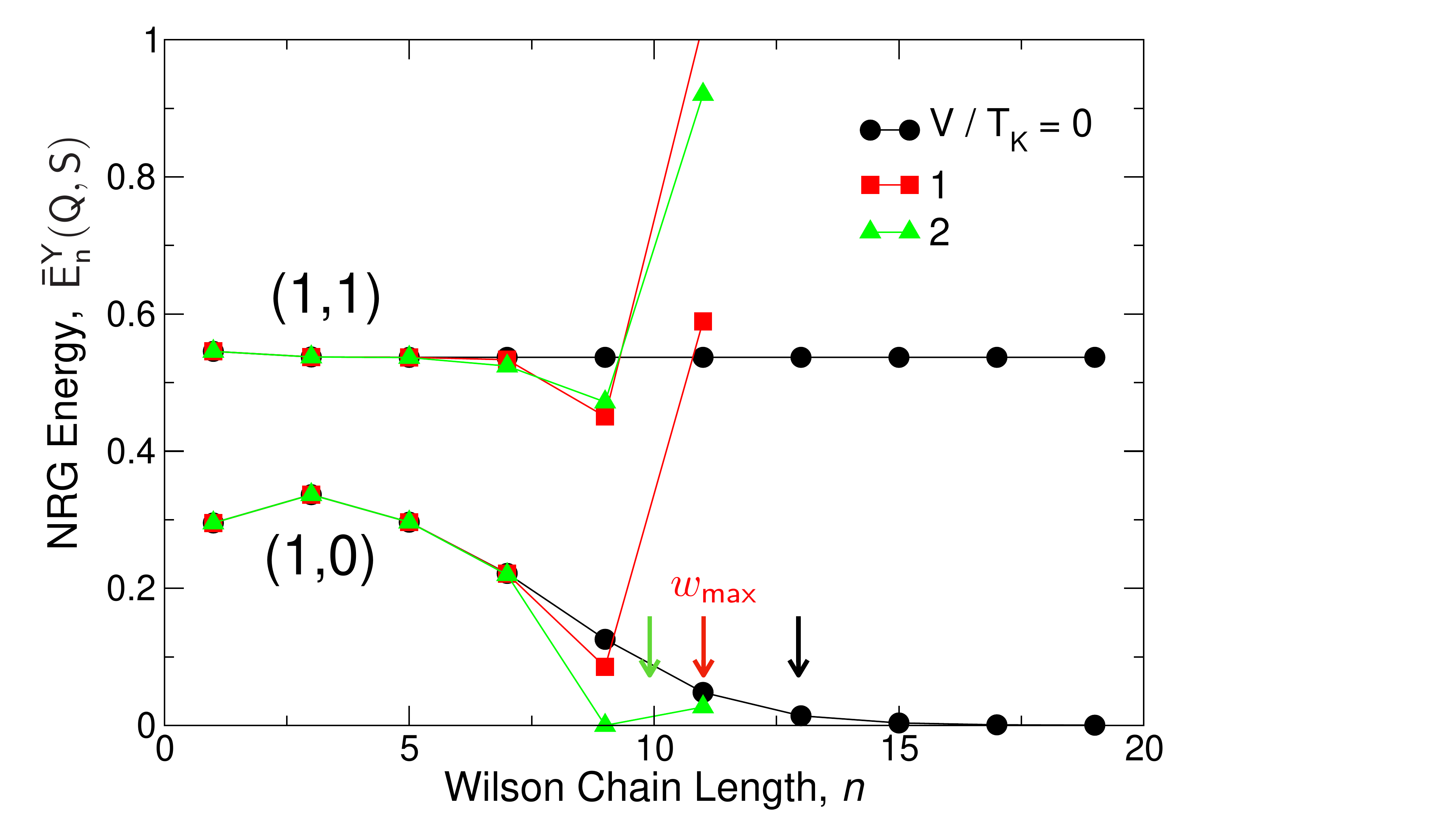}}}
\caption{NRG flow of eigenvalues of $(\hat{H}-V\hat{Y})/D_n$ per $(Q,S)$ at (odd
integer) chain lengths $n$, $\bar{E}^Y_n(Q,S)$, with bias up to $V/2\sim T_K$.
$Q$ and $S$ are charge and spin quantum numbers from the equilibrium ground
state, and $D_n$ is the RG energy scale, Eq.~(\ref{eq:dn}). The
$(1,0)$ sector, Kondo-screened doubly-occupied impurity, approaches the Kondo
fixed point at $V=0$. As the bias grows, the RG flow departs from the
fixed point at the chain length (indicated by an arrow for each bias with
corresponding color that gives the maximum density-matrix weight).
} 
\label{fig3} \end{center}
\end{figure}

Fig.~\ref{fig3} shows the RG flow of the eigenvalues $E^Y_n$ of the (statistical) energy $\hat{H}-V\hat{Y}$ of the WC length $n$, renormalized to the scale $D_n$. The model parameters are $U=1$, $\Gamma=0.1$, and in the particle-hole symmetric limit, $\epsilon_d=0$. The NRG parameters are as follows throughout this work unless mentioned otherwise: $\Lambda=4$, the number of kept states $N_{\rm kept}=\textcolor{black}{2000}$ with the truncation based on the spectrum of $\hat{H}$, the number of $z$-averaging $N_z=8$~\cite{oliveiraPRB1994}. With $L$ and $R$ chains, each NRG iteration has the Hilbert space of dimension $16\times N_{\rm kept}$ on which $\hat{H}$ and $\hat{\rho}_{\rm neq}$ are computed. The total charge and spin $(Q,S)$ defined from the equilibrium ground state are conserved quantum numbers. At zero bias $V=0$, the first excited state $(1,0)$ of spin-singlet ($S=0$) and singly-charged ($Q=1$) configuration approaches the strong-coupling fixed-point as $n\to\infty$~\cite{KWWPRB1980}. As the bias is applied, the flow deviates from the fixed-point starting at $n\approx M$. The $M$ values for the maximum weight $w_{\rm max}$ are marked by arrows for each bias. The calculation shows that physical observables have the dominant contribution when the system is about to depart from the fixed-point. See SM~\cite{SM} for more detailed discussions on the density-matrix weight. \textcolor{black}{The eigenvalues shown in Fig.~\ref{fig3} are measured from the minimum eigenvalue of $(\hat{H}-V\hat{Y})/D_n$ for each WC length $n$, and the overall shift of the spectra by $(E^Y_n)_{\rm min}$ should be added back into $\hat{H}-V\hat{Y}$ for the density-matrix weight.}

\begin{figure}
\begin{center}
\rotatebox{0}{\resizebox{3.4in}{!}{\includegraphics{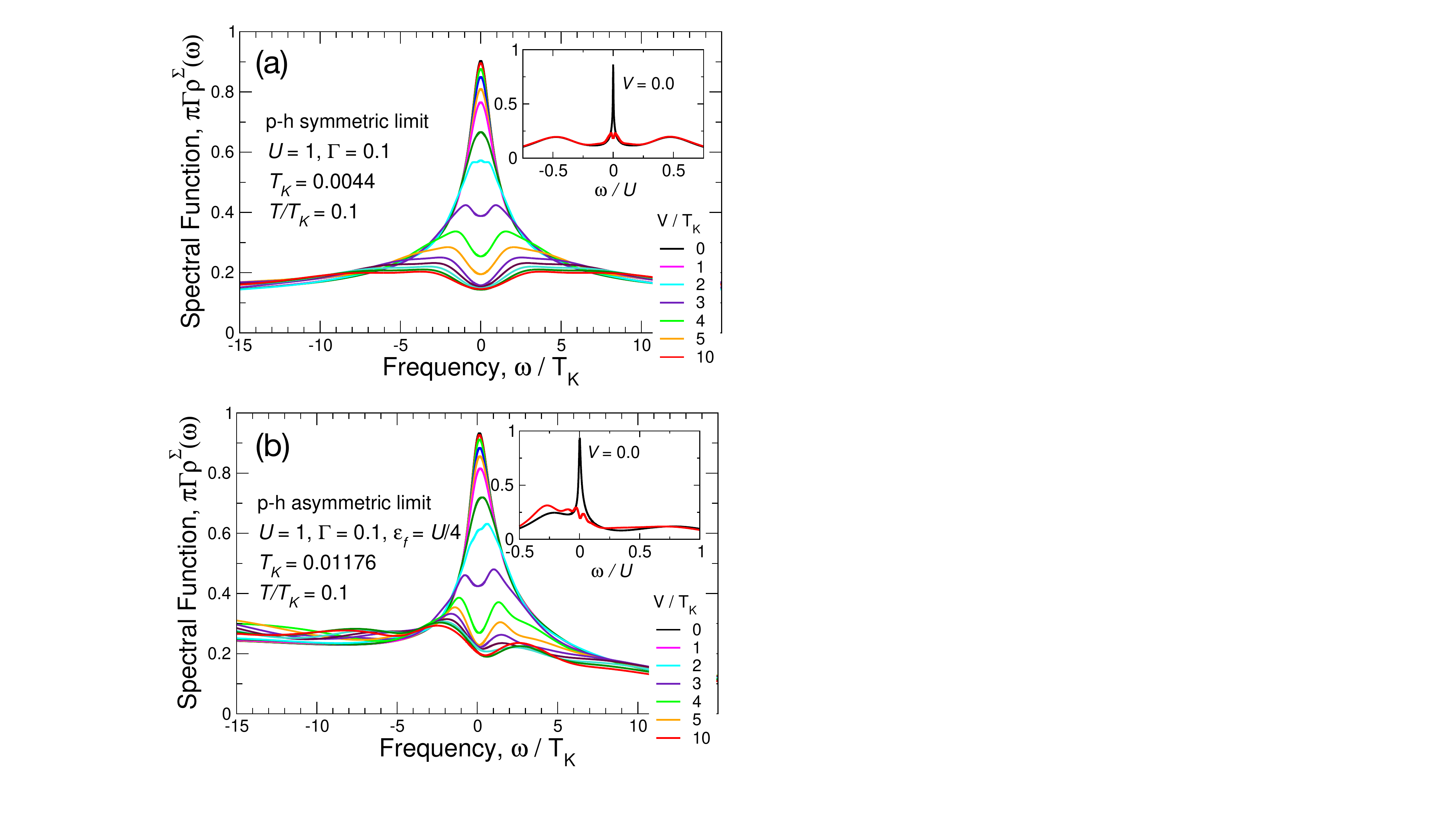}}}
\caption{Spectral evolution of Anderson model under bias $V$.
(a) Particle-hole (p-h) symmetric limit with $\epsilon_d=0$ in the strong Kondo regime ($U/\Gamma=10$, $\Gamma=\Gamma_L+\Gamma_R$) at a temperature much lower than the Kondo temperature $T_K=0.0044$. The right inset shows the whole frequency range demonstrating the sharpness of the Kondo resonance. \textcolor{black}{From top to bottom, $V/T_K=0, 0.2,0.4,0.6, 0.8, 1,2,3,4,5,6,7,8,9,10$ with selected bias values shown in the legend. The peak-splitting begins at $V/2\approx T_K$ with the peak positions at $\omega\approx \pm\frac12 V$.} (b) p-h asymmetric limit $(\epsilon_d=U/4)$ with the same set of bias $V$.
}
\label{fig4}
\end{center}
\end{figure}

We now turn to a detailed discussion of spectral evolution under bias. Calculations for the spectral function calculation follow conventional procedures as  the equilibrium FDM-NRG with the density-matrix given by Eqs.~(\ref{eq:rho}) and (\ref{eq:y}). In practice, due to the rapid growth of the Hilbert space with two WCs, the number of kept states $N_{\rm kept}$ is limited and it has to be carefully checked that we cover the bias energy window of $(-V/2,V/2)$ sufficiently. \textcolor{black}{At the iterations of the maximum FDM weight $w_{\rm max}$, $E_{\rm max}/(V/2)\sim 3-20$ with the maximum energy $E_{\rm max}$ for each iteration of the WS (See SM~\cite{SM} for more details). With $N_{\rm kept}=2000$, the results are well-converged.} We used the usual log-gaussian spectral summation~\cite{sakaiJPSJ1989,bullaRMP2008} with the broadening parameter of $0.44$, after which a uniform gaussian smoothing is performed~\cite{weichselbaumPRL2007} with the width $\delta=0.2\times{\rm max}(T_K,T,V/2)$.  The Kondo temperature $T_K$ is estimated~\cite{KWW2,hewson1997kondo} as $T_K=\sqrt{U\Gamma/2}\,e^{-1/J}$ with $J=(2/\pi)\Gamma(|\epsilon_d-U/2|^{-1}+|\epsilon_d+U/2|^{-1})$.

Fig.~\ref{fig4}(a) presents spectral functions of the particle-hole (p-h) symmetric Anderson model under bias in the strong Kondo regime $U/\Gamma=10$ with $T_K=0.0044$, that is, $T_K\ll\Gamma\ll U$ ($\Gamma/T_K\approx 23$, $U/T_K\approx 227$). The spectral functions are computed by using the so-called self-energy trick~\cite{bullaJPC1998}, denoted as $\rho^\Sigma(\omega)$. It is well-known~\cite{bullaRMP2008} that $\rho^\Sigma(\omega)$ is numerically more reliable than that from the direct spectral summation $\rho(\omega)$. The right-inset in (a) illustrates the sharpness of the Kondo resonance at zero bias compared to the charge peaks at $\omega=\pm U/2$. The main plot in (a) shows the spectral evolution of the Kondo resonance for bias up to $V/T_K=10$. The spectrum at zero bias (black) has the resonance at $\rho^\Sigma(0)\approx 1/(\pi\Gamma)$, satisfying the unitary limit, and its width well matching the analytic $T_K$~\cite{KWW2} as discussed above. In the low-temperature limit ($T=0.1\,T_K$), the Kondo resonance is only slightly affected at low bias $V\ll T_K$. When the (half) bias matches $T_K$ ($V/2\approx T_K$), the Kondo resonance starts to show peak-splitting with a dip at $\omega=0$. Departure from the p-h symmetric limit ($\epsilon_d=U/4$), as shown in (b), does not lead to a significantly different behavior from (a). 

As $V$ increases further, the peak becomes decorrelated and the position aligns well at $\pm V/2$. Although those peaks are often referred to as split Kondo peaks, their linewidth is much greater than $T_K$ and they should be considered weakly correlated.

\textcolor{black}{A good match between $\rho^\Sigma(\omega)$ and $\rho(\omega)$ indicates an internally consistent NRG, and it becomes better for $\eta\ll T_K$. Therefore, we set in the calculations
\begin{equation}
   \eta=0.1\times{\rm max}(T_K,T,V/2)
   \label{eq:eta}
\end{equation}
with $\eta\ll T_K$ for $T,V\ll T_K$, in the similar fashion as discussed in Fig.~\ref{fig2}. This points to the fact that the time-evolution should be taken much longer than the inverse of the emergent energy scale to describe correlated nonequilibrium. With $\eta> T_K$, $\rho^\Sigma(\omega)$ tends to overestimate the spectral weight near $\omega=0$.}

\begin{figure}
\begin{center}
\rotatebox{0}{\resizebox{3.4in}{!}{\includegraphics{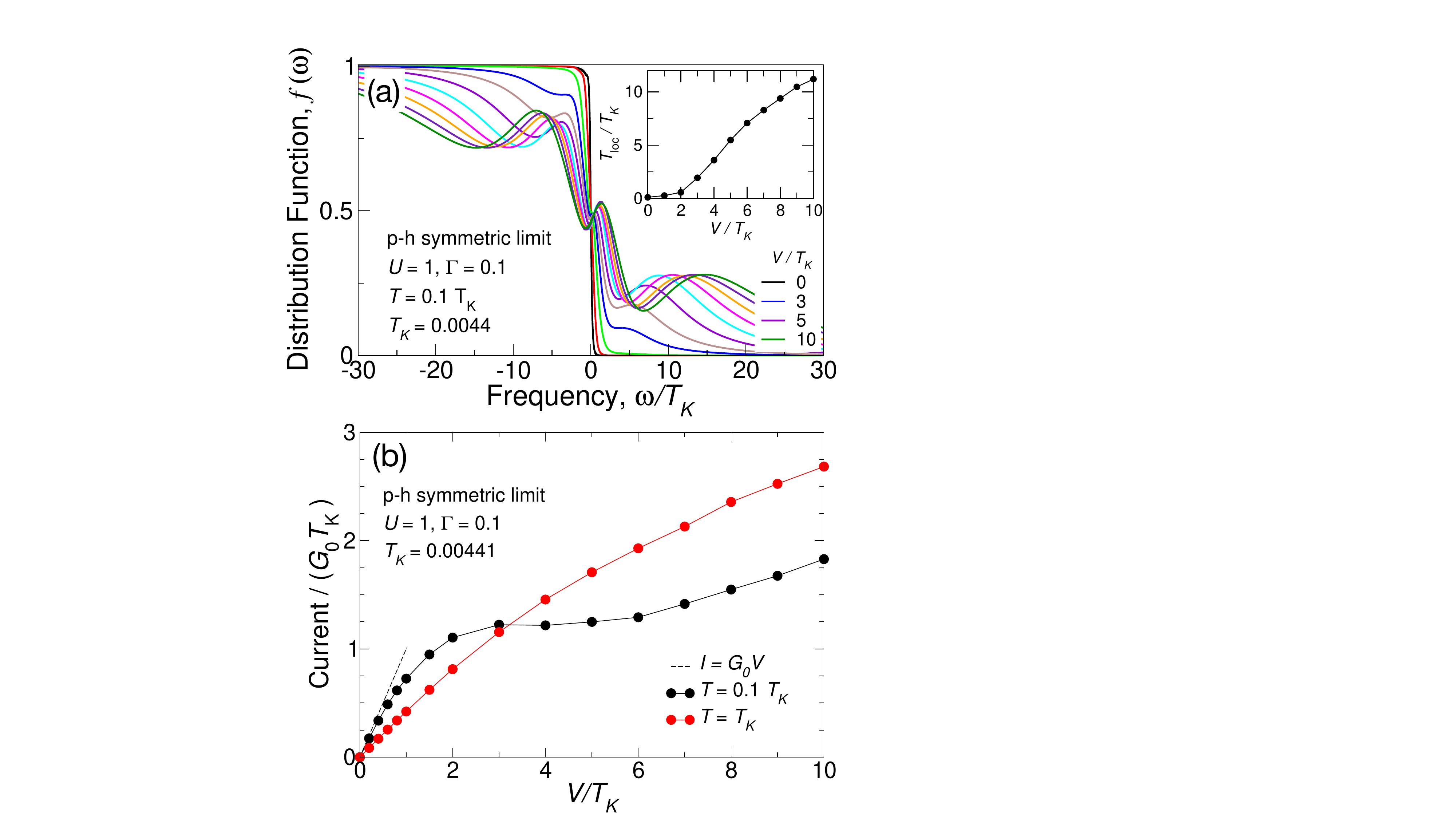}}}
\caption{
(a) \textcolor{black}{The distribution function $f(\omega)$ in the p-h symmetric limit with $U/\Gamma=10$ at bias values $V=0,\cdots,10$ with the spacing $\Delta V=1$. $f(\omega)$ strongly deviates from the Fermi-Dirac function, displaying multiscale features of population inversion at $\omega\sim T_K$ and the broad excitation peaks at $\omega\gtrsim V$. The inset shows the nonequilibrium electron temperature $T_{\rm loc}$ local to the Kondo dot. In the high bias limit, $T_{\rm loc}\approx V$. (b) $IV$ relation at temperatures $T=0.1\,T_K$ and $T_K$. In the low temperature limit at $T=0.1\, T_K$, the unitary conductance limit in the small field limit starts to deviate and reach the current saturation regime for $V\gtrsim T_K$. At higher temperature $T=T_K$, no sign of current saturation is observed.}
}
\label{fig5}
\end{center}
\end{figure}

\textcolor{black}{The statistical evolution reveals richer structures than those of the spectral functions, leading to unusual transport properties, as shown in Fig.~\ref{fig5}. The distribution function, $f(\omega)=G^<_d(\omega)/[2\pi i\rho(\omega)]$ with the $d$-level lesser Green's function $G^<_d(\omega)$, starts as the Fermi-Dirac function at zero bias (black). Starting from $V/2 \approx T_K$, the shape of $f(\omega)$ deviates strongly from the Fermi-Dirac function by developing a large and broad peak at $|\omega|\gtrsim V$. Simultaneously, in the frequency between the split Kondo peaks ($-\frac{V}{2} \lesssim \omega \lesssim \frac{V}{2}$) population inversion occurs with $f^\prime(0)<0$. This population inversion can be understood in terms of the dissociation of the Kondo resonance, with the positive frequency peak more strongly associated with the source ($L$), leading to excess electron occupation and vice versa. It is surprising, however, that the nonequilibrium excitations lead to a very strong particle distribution at frequencies well beyond $V$.
}

\textcolor{black}{The multi-scale distribution is summarized in $T_{\rm loc}$, the nonequilibrium temperature local to the $d$-orbital as
\begin{equation}
    T_{\rm loc}^2=\frac{6}{\pi^2}\int_{-\infty}^\infty\omega[f(\omega)-\Theta(-\omega)]d\omega.
\end{equation}
This Sommerfeld-like definition of the nonequilibrium temperature~\cite{nanolett} has worked as a good indicator in the resistive-switching analysis. $T_{\rm loc}$ in inset of Fig.~\ref{fig5}(a) shows a slow rise in the linear regime, followed by a rapid increase after $V/2\approx T_K$, leading to
\begin{equation}
    T_{\rm loc}\approx V\mbox{ for }\frac{V}{2}\gg T_K.
\end{equation}
}

\textcolor{black}{
 Once the spectral and distribution functions are known, the electric current can be evaluated through the diagrammatic techniques~\cite{meirPRL1992}. Fig.~\ref{fig5}(b) shows the $IV$ relations at low ($T=0.1\,T_K$) and high ($T=T_K$) temperature limit. In the low-$T$ limit, the linear conductance satisfies the unitary limit $G_0=(2\pi)^{-1}$ for $V\ll T_K$. The conductance begins to deviate as the bias increases and for $V/2\gtrsim T_K$ the current begins to saturate until $V\approx 6T_K$. The current saturation behavior agrees with both theoretical~\cite{BolechPRB2016} and the recent numerical~\cite{jeannin2025} predictions. As the temperature increases, as early as $T=T_K$, the current saturation behavior completely vanishes, demonstrating that the current saturation results from strongly correlated effects. At high bias $V>4T_K$, the current is larger in the large temperature limit due to the enhancement by the thermal excitation.   
}


\section{Conclusion}

\textcolor{black}{We represented the nonequilibrium steady-state of the Anderson impurity model with reservoirs at different chemical potentials by formulating Hershfield's scattering-state idea through the numerical renormalization group (NRG) method. By constructing the density-matrix directly using the original basis of the Hamiltonian, we simplify the formulation that could be well-suited for extended models. The results confirm the Kondo peak splitting at bias $V$ larger than the Kondo energy scale $T_K$. The nonequilibrium distribution reveals multiple energy scales with population inversion in the sub-Kondo regime, and strong nonequilibrium excitations well beyond the bias energy. The transport calculation verifies the current saturation for intermediate bias region of $T_K\lesssim \frac{V}{2}\lesssim 3T_K$ in the low temperature limit $T\ll T_K$.}

\textcolor{black}{The proposed method is based on the well-established NRG formulations with the density-matrix replaced by the one built on the statical operator $\hat{H}-V\hat{Y}$. With the implementation of $\hat{Y}$ by using the original basis of the bath states, the method is well-suited for an extension to complex impurity models in nonequilibrium, or for a nonequilibrium impurity solver in lattice models through the dynamical mean-field theory.}

\begin{acknowledgements}
The author thanks F. Anders, N. Andrei, E. Arrigoni, C. J. Bolech, P. Coleman, H. Fotso, G. Kotliar, and A. Weichselbaum for helpful discussions. Computational support from the CCR at the University at Buffalo is acknowledged.
\end{acknowledgements}

\bibliography{nrglib.bib}

\end{document}


\myfonts

\centerline{\textsf{\Large Supplementary Material: Nonequilibrium Spectral Distribution of Biased Kondo Resonance}
}
\bigskip
\centerline{\sf\large Jong E. Han}

\vspace{1cm}

\noindent\textbf{A. Density-Matrix Weight $w_n$ with the Wilson-Chain Length $n$, Computational Cost}


\begin{figure}[h]
\rotatebox{0}{\resizebox{6.0in}{!}{\includegraphics{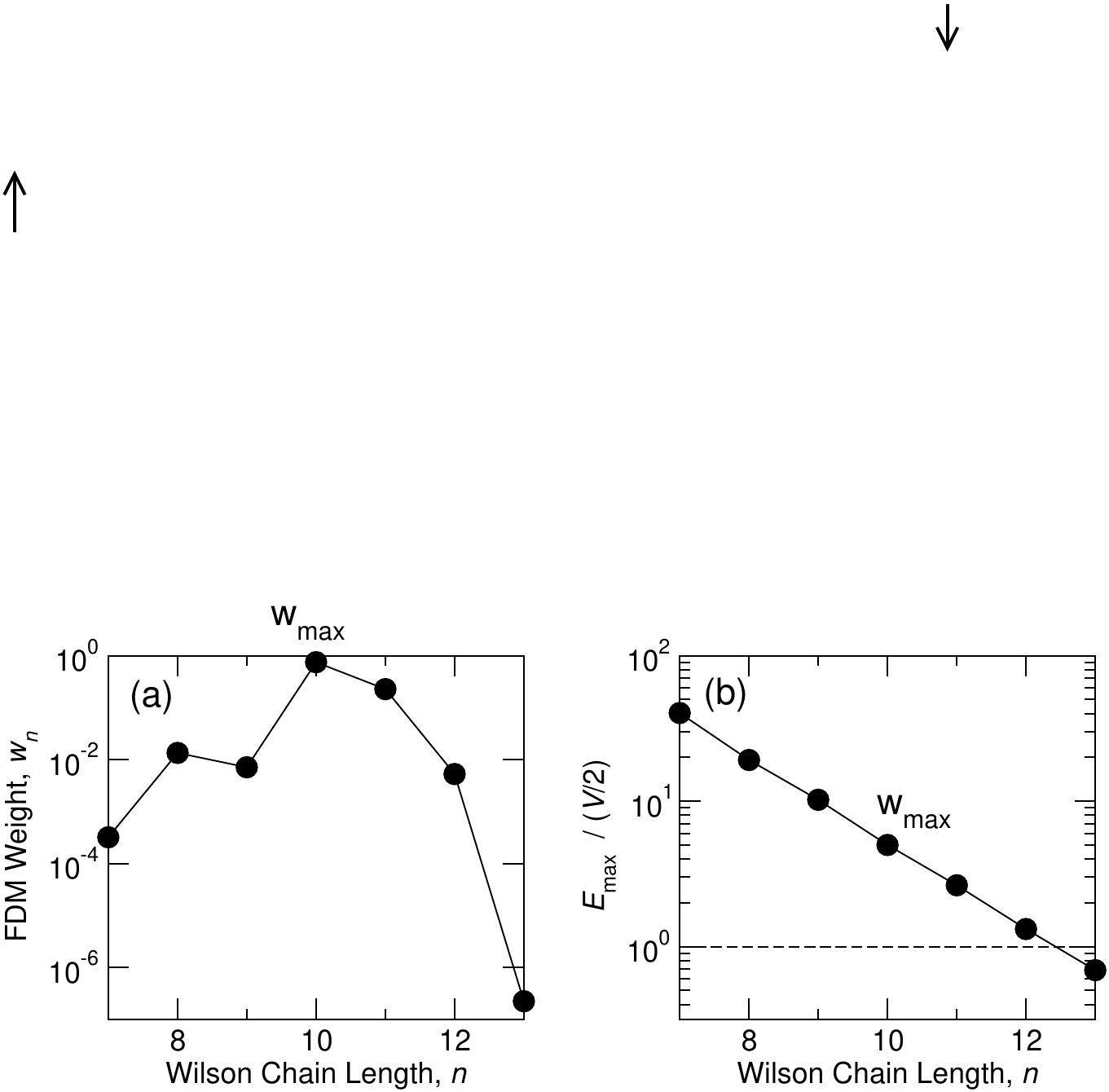}}}
\caption{(a) Full-density-matrix (FDM) weight $w_n$ as a function of the Wilson chain length $n$ for $U/\Gamma=10$, $T/T_K=0.1$, $V/T_K=3$. The weight is peaked at $n=10$. (b) The maximum energy $E_{\rm max}$ vs the bias $V/2$ as a function of the Wilson-chain length $n$. $E_{\rm max}$ is defined in the energy scale of the unrenormalized Hamiltonian $\hat{H}$. It is desirable to have $E_{\rm max}/(V/2)>1$ to cover the excited energies, which is satisfied for dominant density-matrix.
}
\label{figS2}
\end{figure}






Apart from using the density-matrix Eq.~(2) of the main text for the nonequilibrium full-density-matrix numerical renormalization group (FDM-NRG), the structure of the NRG algorithm is the same as in previous works~\cite{weichselbaum}. The behavior of the density matrix weight and the range of energy eigenvalues as a function of the NRG chain is illustrated in Fig.~S1. The weight of the Wilson shell reaches a maximum over a few $n$ values, in this case $n=10$. Usually $w_n$ drops precipitously after the maximum, and we may truncate further NRG iterations. In (b), we monitor the energy range that observables are recorded. The maximum energy window $E_{\rm max}$ for updating observables should exceed the bias window $V/2$. In the dominant Wilson shell $n=8-11$, $E_{\rm max}/(V/2) \gtrsim 3$.

In practice, the NRG iterations are repeated three times: (i) to identify the maximum weight $w_n$ of the density matrix for the Wilson chain length $n$, (ii) to produce the basis transformation matrix~\cite{weichselbaum,hofstetter} to compute the reduced density-matrix, (iii) for a production run to compute observables, such as the spectral weight. With the size of the Hilbert space $N_{\rm sp}$ at each NRG iteration being $N_{\rm sp}=16 N_{\rm kept}$ ($16$ for the two added sites to the $L$- and $R$-Wilson chains, $N_{\rm kept}$ for the number of kept states from the previous iteration), the computational times for the diagonalization of the Hamiltonian $\hat{H}$, the construction of the $Y$-operator and the exponentiation of $\hat{H}-V\hat{Y}$, and the computation of the spectral coefficients scale as $N_{\rm sp}^3$. Each stages have similar computational times. 

With Intel-Xeon-Gold 6330 processors, computations with $N_{\rm kept}=2000$ and the maximum Wilson chain length of 10 took roughly 60 minutes with the memory size of 15 GB per each $z$ value. Computations for multiple $z$ values can easily be parallelized.

\bigskip

\noindent\textbf{B. Fortran Code for $Y$-operator in the Non-interacting Resonant-Level Model}

\bigskip

The Fortran code to generate Fig.~2 in the main text (without the $z$-averaging) is shown. The flow of the code is summarized as follows. Definition of symbols is given in the main text.
\begin{enumerate}[(1)]
    \item Define the hopping parameters (with $z=0$) as~\cite{bullaRMP2008}
    \begin{equation}
        t_n=\frac{(1+\Lambda^{-1})(1-\Lambda^{-n-1})}{2\sqrt{1-\Lambda^{-2n-1}}\sqrt{1-\Lambda^{-2n-3}}}\Lambda^{-n/2}.
    \end{equation}
    \item Set up the Hamiltonian matrix for the tight-binding chain including the central $d$-site with the one-particle basis $|i\rangle$ ($i=$ site index), and obtain the eigenvalues $E_\alpha$ with single-particle eigenvectors $|\alpha\rangle$
    \item Set up the number operators $\langle\alpha|N_{L/R}|\beta\rangle=\sum_{i\in L/R}\langle\alpha|i\rangle\langle i|\beta\rangle$ and the initial $Y$-operator $\hat{Y}_0=\frac12(\hat{N}_L-\hat{N}_R)$.
    \item Define the current operator $\langle\alpha|\hat{J}|\beta\rangle=\sum_{ij}\langle\alpha|i\rangle\langle i|\hat{J}|j\rangle\langle j|\beta\rangle$.
    \item Loop for voltage values $V$
    \begin{enumerate}[(i)]
    \item Define $\eta={\rm min}(\eta_0\sqrt{\Gamma^2+V^2},\Gamma)$.
    \item Construct the full $Y$-operator by
    \begin{equation}
        \langle\alpha|\hat{Y}|\beta\rangle=\frac{i\eta}{-E_\alpha+E_\beta+i\eta}\langle\alpha|\hat{Y}_0|\beta\rangle.
    \end{equation}
    \item Obtain the Hershfield's density matrix $\hat\rho_{\rm neq}=\exp[-(\hat{H}-V\hat{Y})/T]/Z_{\rm neq}$, first  by diagonalizing $\hat{H}-V\hat{Y}$ with eigenvalues $\epsilon_\lambda$ and eigenvectors $|\lambda\rangle$, and forming $\langle\alpha|\hat\rho_{\rm neq}|\beta\rangle=\sum_\lambda \langle\alpha|\lambda\rangle (1+e^{\epsilon_\lambda/T})^{-1}\langle\lambda|\beta\rangle$ 
    \item Compute the current expectation value
    $\langle\hat{J}\rangle=\sum_{\alpha\beta}\langle\alpha|\hat{J}|\beta\rangle\langle\beta|\hat\rho_{\rm neq}|\alpha\rangle.$
    \end{enumerate}
\end{enumerate}
For $z$-averaging, $t_n$ should be modified at non-zero $z$ values as well-known in the literature~\cite{bullaRMP2008}, and take the average of $\langle\hat{J}\rangle$ over $z$-values.

\begin{verbatim}
module types
   implicit none
   integer, parameter  ::  i4b=selected_int_kind(9)
   integer, parameter  ::  dp=kind(1.0d0)
end module types

module globals
   use types
   implicit none
   integer(i4b), parameter :: msite=100,max=2*msite+1
   real(dp) :: lambda,V,temp,D0,t0
   real(dp) :: tn(-1:max),eta,current,g0,ith,h(max,max),eig(max),evec(max,max), &
            en1(max),work(20*max),eta0,gamma
   complex*16 :: yop(max,max),y0op(max,max),hy(max,max),zvec(max,max),zwork(20*max), &
            rhoneq(max,max),cij(max,max)
   real(dp), parameter :: pi=2.d0*dasin(1.d0)
   complex*16, parameter :: ii=(0.d0,1.d0)
end module globals
!
!	===============================================
!	Non-interacting resonant level calculation
!	===============================================
!
program NIRLM

   use types
   use globals
   implicit none
   integer(i4b) :: i,j,k,iv,info,l
   real(dp) :: sum,yR,yL
   complex*16 zsum
   
   open(5,file='input.par')
   open(16,file='curr.dat')
!
!	--------------------------------------------------
!	Wilson chain parameters
!	--------------------------------------------------
!
   read(5,*)
   read(5,*) D0,lambda,t0,temp,eta0
!
   do l=0,msite-1
   
      tn(l) = 0.5d0*(1.0d0+1.0/lambda)/sqrt(lambda**(l-1))* &
              (1.0d0-lambda**(-l-1))* &
              ((1.0d0-lambda**(-2*l-1))*(1.0d0-lambda**(-2*l-3)))**(-0.5d0)
      write(6,*) l,tn(l)
   
   enddo

   h = 0.0
!
!	--------------------------------------------------
!	Hopping in Wilson chain (WC)
!	--------------------------------------------------
!
   do l=0,msite-2
      i = l + msite + 2
      h(i,i+1) = -tn(l)
      h(i+1,i) = -tn(l)
      i = -l + msite
      h(i,i-1) = -tn(l)
      h(i-1,i) = -tn(l)
   enddo
!
!	--------------------------------------------------
!	Hopping between WC and the dot
!	--------------------------------------------------
!
   h(msite+1,msite+2) = -t0
   h(msite+2,msite+1) = -t0
   h(msite+1,msite) = -t0
   h(msite,msite+1) = -t0
!
!	--------------------------------------------------
!	BLAS-LAPACK routine for diagonalization
!	--------------------------------------------------
!
   evec = h
   call dsyev('V','U',max,evec,max,en1,work,20*max,info)
!
!	--------------------------------------------------
!	Y_0 operator
!	--------------------------------------------------
!
   do i=1,max
      do j=1,max
         yR = 0.0
         do k=msite+2,2*msite+1
            yR = yR + evec(k,i)*evec(k,j)
         enddo
         yL = 0.0
         do k=1,msite
            yL = yL + evec(k,i)*evec(k,j)
         enddo
         y0op(i,j) = 0.5*(yL-yR)
      enddo
   enddo
!
!	--------------------------------------------------
!	current operator in the basis of H
!	--------------------------------------------------
!
   do i=1,max
   do j=1,max
      zsum = evec(msite+1,i)*evec(msite+2,j)
      zsum = zsum - evec(msite+2,i)*evec(msite+1,j)
      zsum = zsum - evec(msite+1,i)*evec(msite,j)
      zsum = zsum + evec(msite,i)*evec(msite+1,j)
      cij(i,j) = -0.5*ii*t0*zsum
   enddo
   enddo
!
!	--------------------------------------------------
!	Start of bias sweep
!	--------------------------------------------------
!
   do iv=0,400

      V = 0.0001*iv
      eta = dmin1(gamma,eta0*sqrt(gamma**2+V**2))
!
!	--------------------------------------------------
!	full Y operator, Eq. (5)
!	--------------------------------------------------
!
      do i=1,max
         do j=1,max
            yop(i,j) = ii*eta/(-en1(i)+en1(j)+ii*eta)*y0op(i,j)
         enddo
      enddo
!
!	--------------------------------------------------
!	define H-VY
!	--------------------------------------------------
!
      hy = 0.0
      do i=1,max
         hy(i,i) = en1(i)
      enddo
      hy = hy - V*yop

      zvec = hy

      call zheev('V','U',max,zvec,max,eig,zwork,20*max,work,info)
!
!	--------------------------------------------------
!	exponentiation of H-Y for the density matrix in the basis of H
!	--------------------------------------------------
!
      do i=1,max
      do j=1,max
         zsum = 0.0
         do k=1,max
            zsum = zsum + zvec(i,k)*dconjg(zvec(j,k))/(1.0+exp(eig(k)/temp))
         enddo
         rhoneq(i,j) = zsum
      enddo
      enddo
!
!	-----------------------------------------------------------
!	trace of (current*rho) for the current
!	ith is the exact theoretical answer
!	-----------------------------------------------------------
!
      g0 = pi*t0**2/(2.0*D0)
      ith = 4.0*g0*datan(V/(4.0*g0))/(2.0*pi)
      current = 0.0
      do i=1,max
         do j=1,max
            current = current + rhoneq(i,j)*cij(j,i)
         enddo
      enddo
      write(16,'(3f20.12)') V,2.0*pi*current,2.0*pi*ith

   enddo
   
end program NIRLM
\end{verbatim}